\documentclass[10pt,showpacs,amsmath,amssymb]{article}
\usepackage{amsfonts}
\usepackage{mathdots}
\usepackage{color}

\begin{document}

\title{General solutions of the supersymmetric $\mathbb{C}P^2$ sigma model and its generalisation to $\mathbb{C}P^{N-1}$}


\author{Laurent Delisle, V\'eronique Hussin and Wojtek J. Zakrzewski}

\author{L. Delisle${}^{1,5}$, V. Hussin${}^{2,3,6}$ and W. J. Zakrzewski${}^{4,7}$}

\footnotetext[1]{Institut de math\'ematiques de Jussieu-Paris Rive Gauche, UP7D-Campus des Grands Moulins, B\^atiment Sophie Germain, Cases 7012, 75205 Paris Cedex 13.}
\footnotetext[2]{D\'epartement de Math\'ematiques et de
Statistique, Universit\'e de Montr\'eal, C.P. 6128,
Succ.~Centre-ville, Montr\'eal (Qu\'ebec) H3C 3J7, Canada.}
\footnotetext[3]{Centre de Recherches Math\'ematiques,
Universit\'e de Montr\'eal, C.P. 6128, Succ.~Centre-ville,
Montr\'eal (Qu\'ebec) H3C 3J7, Canada.}
\footnotetext[4]{Department of Mathematical Sciences, University of Durham, Durham DH1 3LE, United Kingdom.}
\footnotetext[5]{email:laurent.delisle@imj-prg.fr}
\footnotetext[6]{email:hussin@dms.umontreal.ca}
\footnotetext[7]{email:w.j.zakrzewski@durham.ac.uk}

\date{\today}

\maketitle

\begin{abstract}
A new approach for the construction of finite action solutions of the supersymmetric $\mathbb{C}P^{N-1}$ sigma model is presented. We show that this approach produces more non-holomorphic solutions than those obtained in previous approaches.  We study the $\mathbb{C}P^2$ model in detail and present its solutions in an explicit form. We also show how to generalise this construction to $N>3$.
\end{abstract}


\section{Introduction}
Study of exact solutions of integrable models is a subject of great interest to the mathematics and physics communities. In addition, the 2-dimensional  integrable bosonic $\mathbb{C}P^{N-1}$ sigma model has found applications in physics, biology, and mathematics \cite{gross,safram,davydov,rajaraman,manton,Landolfi}. The solutions of this model are known \cite{wojtekbook} and were used to construct solutions of more general grassmannian sigma models \cite{wojtekbook} and to generate surfaces immersed in the Lie algebra $su(N)$ \cite{delisle1,delisle2,delisle3,grundland1,grundland,hussin,bolton,jiao,jiao1,jie,peng,peng1}
. Indeed, the present authors have discussed the solutions of general grassmannian models which correspond to  surfaces of constant gaussian curvatures \cite{delisle1,delisle2,delisle3}. Other papers discussed also different geometric quantities of these surfaces such as their fundamental forms, their mean curvatures and the Willmore functional \cite{delisle3,grundland1,grundland}.

In a recent paper \cite{delisle}, we studied the surfaces obtained from the solutions of the supersymmetric $\mathbb{C}P^{N-1}$ sigma models. We classified the surfaces of constant gaussian curvature associated to holomorphic solutions and showed their deep connection with the bosonic Veronese curve. This curve was shown to be useful to obtain a classification of all constant curvature surfaces in the bosonic $\mathbb{C}P^{N-1}$ model when one generates the complete set of solutions of this model via the repeated application of an orthogonalisation operator \cite{bolton}. 

In this same paper \cite{delisle}, we obtained some non-holomorphic solutions using two different strategies: the first one relied on imposing supersymmetric invariance and, the other, on forcing the conformality of the constructed surfaces. Although we managed to present some non-holomorphic solutions, a general method of constructing all of them  still has not been found and the present paper addresses this problem. Our main idea here is to generalize the so-called holomorphic method \cite{wojtekbook}, used in constructing the complete set of solutions of the bosonic model, and to apply it to the supersymmetric $\mathbb{C}P^{N-1}$ sigma model. So, let us remind the reader of the bosonic holomorphic method. In this method we consider a holomorphic $N$-component vector $\psi_0=\psi_0(x_+)$ such that the sequence 
\begin{equation}
\psi_0,\quad \psi_1=\partial_+\psi_0,\quad\psi_2=\partial_+\psi_1,\cdots,\quad\psi_{N-1}=\partial_+\psi_{N-2}\label{vectbos}
\end{equation}
forms a linearly independent set of vectors. Using Gram-Schmidt, we orthogonalize the vectors of the above set and, then, obtain a new set consisting of $N$ vectors
\begin{equation}
z_0=\psi_0,\quad z_j=\psi_j-\sum_{k=0}^{j-1}\frac{z_k^{\dagger}\psi_j}{\vert z_k\vert^2}z_k,\quad j=1,2,\cdots,N-1.
\end{equation}
Then, as is well known \cite{wojtekbook}, the vectors 
\begin{equation}
Z_j=\frac{z_j}{\vert z_j\vert},\quad j=0,1,\cdots,N-1,
\end{equation}
are solutions of the Euler-Lagrange equations of the model. Furthermore, it is also known that the set $\{Z_0,Z_1,\cdots,Z_{N-1}\}$ is complete and thus all solutions are of this type. In our recent paper \cite{delisle}, we have asked the question whether a similar classification of solutions existed in the supersymmetric case. In that paper, we were not able to present a definitive answer to this question. There exists other papers which have also looked for solutions of this model. In \cite{fujii}, the authors solved the linear Dirac equation associated to the model. This reduced model only considered quadratic fermionic terms in the Lagrangian eliminating higher order odd terms. In another paper \cite{din}, the authors considered the fermionic contributions as commuting quantities. 
This method has given formal solutions but some practical uncertainties as to the validity of this procedure had been  raised. In the  present paper, we go a step further and present a systematic method for constructing solutions of the general model without any of the above assumptions. In fact, our approach is based on a generalisation of the holomorphic method for constructing solutions in the bosonic model \cite{wojtekbook}.

In order to make the paper self-contained, we present, in the next section, a brief description of the supersymmetric $\mathbb{C}P^{N-1}$ sigma model and of its formulation in terms of orthogonal projectors. We then discuss, in detail, the $\mathbb{C}P^{2}$ model, the simplest one possessing  non-holomorphic solutions, and we  construct its solutions  using our new approach. This approach is a generalization to the supersymmetric context of the general construction in the bosonic case \cite{wojtekbook}. We show that our method enlarges a class of solutions and even though we feel that it generates all of them we have no proof that this is the case. We finish this section by presenting some explicit examples of the obtained solutions. In section 4, we generalize our procedure to the more general supersymmetric $\mathbb{C}P^{N-1}$ sigma models and make some remarks on constraints that have to be imposed in the construction of solutions of these models. We conclude the paper with some remarks and our future outlook.

\section{The supersymmetric $\mathbb{C}P^{N-1}$ sigma model}
In this section, for completeness, we recall the definition of  the two-dimensional $\mathbb{C}P^{N-1}$ supersymmetric sigma model and its orthogonal projector formulation \cite{wojtekbook}. We will use this formulation in the subsequent sections.

The two-dimensional supersymmetric $\mathbb{C}P^{N-1}$ sigma model involves the collection of bosonic superfields $\Phi$ defined on the complex superspace of local coordinates $(x_+,x_-;\theta_+,\theta_-)$ with values in the Grassmannian manifold $\mathbb{C}P^{N-1}$. In this formulation, $(x_+,x_-)$ are local coordinates of the complex plane $\mathbb{C}$ ($x_+^{\dagger}=x_-$) and $(\theta_+,\theta_-)$ are complex odd Grassmann variables satisfying
\begin{equation}
\theta_+\theta_-+\theta_-\theta_+=0,\quad \theta_+^2=\theta_-^2=0,\quad \theta_+^{\dagger}=\theta_-.
\end{equation}
The superfields $\Phi$ are $N$-components vectors and satisfy the nonlinear condition
\begin{equation}
\Phi^{\dagger}\Phi=1.\label{constraint}
\end{equation}
The classical solutions of the model, which we want to discuss, are actually critical points of finite energy of the action functional defined in terms of the Lagrangian density
\begin{equation}
\mathcal{L}=2(\vert\check{D}_+\Phi\vert^2-\vert\check{D}_-\Phi\vert^2),\label{lagrangian}
\end{equation}
where $\check{D}_{\pm}\Lambda=\check{\partial}_{\pm}\Lambda-\Lambda(\Phi^{\dagger}\check{\partial}_{\pm}\Phi)$ are the covariant derivatives constructed from the gauge-invariance of the Lagrangian density under the transformation 
\begin{equation}
\Phi\quad \longrightarrow\quad V\Phi U,
\end{equation} 
where $U\in U(1)$ and $V\in U(N)$ are respectively local and global gauge transformations of the unitary group.
The operators $\check{\partial}_{\pm}$ are odd derivatives defined by
\begin{equation}
\check{\partial}_{\pm}=-i\partial_{\theta_{\pm}}+\theta_{\pm}\partial_{\pm},
\end{equation}
that satisfy
\begin{equation}
{\check{\partial}_{\pm}}^2=-i \partial_{\pm}=-i \partial_{x_{\pm}}.
\end{equation}
We use the following convention: $\vert\Lambda\vert^2=\Lambda^{\dagger}\Lambda$ for $\Lambda$ a bosonic or a fermionic field \cite{cornwell}. This convention is used to eliminate the ambiguity for fermionic constants. As is well known the Euler-Lagrange equations
corresponding to (\ref{lagrangian}) are given by:
\begin{equation}
\check{D}_+\check{D}_-\Phi+\vert\check{D}_-\Phi\vert^2\Phi=0,\quad \Phi^{\dagger}\Phi=1.\label{EulerLagrange}
\end{equation}

As we are interested only in finite energy solutions of (\ref{EulerLagrange}), we have to impose the boundary conditions. We actually want the fields $\Phi$ (or strictly speaking their bosonic part) to converge to a constant at infinity, and sufficiently fast. We thus impose the boundary conditions
\begin{equation}
\check{D}_{\pm}\Phi\longrightarrow0,{\rm as} \quad \vert x_+\vert\longrightarrow\infty.
\end{equation}
These boundary conditions have the effect of compactifying the bosonic part of the superspace $(x_+,x_-;\theta_+,\theta_-)$ into the 2-sphere $S^2$ via the stereographic projection. 



A convenient reformulation of the two-dimensional $\mathbb{C}P^{N-1}$ supersymmetric sigma model involves considering it  in a gauge-invariant way in terms of orthogonal projectors \cite{wojtekbook,delisle,hussin1}. Indeed, let $\Phi$ be a solution of the model and define $\mathbb{P}$ as
\begin{equation}
\mathbb{P}=\Phi\Phi^{\dagger}.
\end{equation}
Then using (7), we see that $\mathbb{P}\longrightarrow V\mathbb{P}V^{\dagger}$ which corresponds to a global gauge transformation, since $V$ is independent of local coordinates. Using the nonlinear constraint (\ref{constraint}), we deduce that $\mathbb{P}$ is a rank-one orthogonal projector, \textit{i.e.} it possesses the following properties
\begin{equation}
\mathbb{P}^2=\mathbb{P}^{\dagger}=\mathbb{P},\quad \hbox{Tr}\,\mathbb{P}=1.
\end{equation}
In this setting, the Lagrangian density (\ref{lagrangian}) takes the form
\begin{equation}
\mathcal{L}=2\hbox{Tr}\left(\check{\partial}_-\mathbb{P}\check{\partial}_+\mathbb{P}\right)
\end{equation}
and the Euler-Lagrange equations (\ref{EulerLagrange}) can be equivalently rewritten as
\begin{equation}
[\check{\partial}_+\check{\partial}_-\mathbb{P},\mathbb{P}]=0,\quad \mathbb{P}^2=\mathbb{P},\label{EL}
\end{equation}
where $[A,B]=AB-BA$ is the usual matrix commutator. Furthermore, a fact that will become useful in the subsequent sections is that the Euler-Lagrange equations (\ref{EL}) may be written as a super-conservation law:
\begin{equation}
\check{\partial}_+\Xi+\check{\partial}_-\Xi^{\dagger}=0,\quad \Xi=[\mathbb{P},\check{\partial}_-\mathbb{P}].\label{susycons}
\end{equation}
This equation differs from the one obtained in the bosonic case \cite{wojtekbook}. Here, the equation involves the odd derivatives $\check{\partial}_{\pm}$ and the fermionic quantities $\Xi$ and $\Xi^{\dagger}$.

\section{The $\mathbb{C}P^2$ model}
In this section, we  consider the simplest model which possesses non-holomorphic solutions and we present a systematic method of construction of all solutions of this model.

\subsection{Solutions of the $\mathbb{C}P^2$ model}
Let us consider, as in (\ref{vectbos}), a holomorphic three component superfield $\tilde{\psi}_0=\tilde{\psi}_0(x_+,\theta_+)$ from which we construct a linearly independent sequence of holomorphic bosonic supervectors $\{\tilde{\psi}_0,\tilde{\psi}_1,\tilde{\psi}_2\}$  from the following expressions
\begin{equation}
\tilde{\psi}_0,\quad \Gamma_1\tilde{\psi}_1=\check{\partial}_+\tilde{\psi}_0,\quad \Gamma_2\tilde{\psi}_2=\check{\partial}_+\tilde{\psi}_1,\label{systemCP2}
\end{equation}
where $\Gamma_1(x_+,\theta_+)$ and $\Gamma_2(x_+,\theta_+)$ are $3\times 3$ fermionic matrices \cite{cornwell}. 
Of course, as $\mathbb{C}P^2\subset \mathbb{C}^3$ the vectors $\tilde{\psi}_i$ can be used as the basis vectors in our construction. Moreover, using  the Gram-Schmidt procedure, we can also construct a set of three bosonic orthogonal supervectors 
\begin{equation}
\tilde{z}_0=\tilde{\psi}_0,\quad \tilde{z}_1=\tilde{\psi}_1-\frac{\tilde{z}_0^{\dagger}\tilde{\psi}_1}{\vert\tilde{z}_0\vert^2}\tilde{z}_0,\quad \tilde{z}_2=\tilde{\psi}_2-\frac{\tilde{z}_1^{\dagger}\tilde{\psi}_2}{\vert\tilde{z}_1\vert^2}\tilde{z}_1-\frac{\tilde{z}_0^{\dagger}\tilde{\psi}_2}{\vert\tilde{z}_0\vert^2}\tilde{z}_0.
\end{equation}
They lead to the orthogonal projectors
\begin{equation}
\mathbb{P}_j=\Phi_i\Phi_i^{\dagger},\quad \hbox{where}\quad \Phi_j=\frac{\tilde{z}_j}{\vert \tilde{z}_j\vert}
,\quad j=0,1,2,\label{solCP2}
\end{equation}
which then  satisfy the completeness relation
\begin{equation}
\mathbb{P}_0+\mathbb{P}_1+\mathbb{P}_2=\mathbb{I}.\label{complete}
\end{equation}

Next, we have to make sure that these projectors are solutions of the Euler-Lagrange equation (\ref{susycons}). In fact, this requirement will imply a constraint on the superfield $\Gamma_1\tilde{\psi}_1$. Then,  with such a constraint, we will find the admissible forms of the  bosonic superfields 
 $\tilde{\psi}_0,\tilde{\psi}_1,\tilde{\psi}_2$ (and so of the corresponding bosonic superfields $\tilde{z}_0$, $\tilde{z}_1$, $\tilde{z}_2$).
 
We thus start by explicitly computing $\Xi=\Xi_0$ in (\ref{susycons}) for the projector $\mathbb{P}_0$. We use the identities
\begin{equation}
\check{\partial}_-\tilde{z}_0=0,\quad \check{\partial}_+\left(\frac{\tilde{z}_0}{\vert\tilde{z}_0\vert^2}\right)=\frac{1}{\vert\tilde{z}_0\vert^2}(\mathbb{I}-\mathbb{P}_0)\check{\partial}_+\tilde{z}_0=\frac{1}{\vert\tilde{z}_0\vert^2}(\mathbb{I}-\mathbb{P}_0)\Gamma_1\tilde{\psi}_1,
\end{equation}
from which it follows that
\begin{equation}
\check{\partial}_+\mathbb{P}_0=\frac{1}{\vert\tilde{z}_0\vert^2}(\mathbb{I}-\mathbb{P}_0)(\Gamma_1\tilde{\psi}_1) \tilde{z}_0^{\dagger}.
\end{equation}
Thus, we have
\begin{equation}
\Xi_0^{\dagger}=[\check{\partial}_+\mathbb{P}_0,\mathbb{P}_0]=\check{\partial}_+\mathbb{P}_0
\end{equation}
which clearly satisfies (\ref{susycons}) since  $\{\check{\partial}_+,\check{\partial}_-\}=0$ and $\mathbb{P}_0^{\dagger}=\mathbb{P}_0$.

The explicit determination of $\Xi=\Xi_1$ in (\ref{susycons}) for the projector $\mathbb{P}_1$ requires more calculations. First, we note that
\begin{equation}
\check{\partial}_-\tilde{z}_1=-(\check{\partial}_-\mathbb{P}_0)\tilde{\psi}_1=-\frac{(\Gamma_1\tilde{\psi}_1)^{\dagger}\tilde{z}_1}{\vert \tilde{z}_0\vert^2}\tilde{z}_0\label{derimoinsz1}
\end{equation}
and, as a consequence of the orthogonality between $\tilde{z}_0$ and $\tilde{z}_1$, we see that
\begin{equation}
\check{\partial}_+\left(\frac{\tilde{z}_1}{\vert\tilde{z}_1\vert^2}\right)=\frac{1}{\vert\tilde{z}_1\vert^2}(\mathbb{I}-\mathbb{P}_1)\check{\partial}_+\tilde{z}_1.
\end{equation}
But, we have that
\begin{equation}
\check{\partial}_+\tilde{z}_1=-(\check{\partial}_+\mathbb{P}_0)\tilde{\psi}_1+(\mathbb{I}-\mathbb{P}_0)\check{\partial}_+\tilde{\psi}_1=(\mathbb{I}-\mathbb{P}_0)\left(\Gamma_2\tilde{\psi}_2-\frac{\tilde{z}_0^{\dagger}\tilde{\psi}_1}{\vert \tilde{z}_0\vert^2}\Gamma_1\tilde{\psi}_1\right),
\end{equation}
from which we deduce the expression
\begin{equation}
\check{\partial}_+\left(\frac{\tilde{z}_1}{\vert\tilde{z}_1\vert^2}\right)=\frac{1}{\vert \tilde{z}_1\vert^2}(\mathbb{I}-\mathbb{P}_0-\mathbb{P}_1)\left(\Gamma_2\tilde{\psi}_2-\frac{\tilde{z}_0^{\dagger}\tilde{\psi}_1}{\vert \tilde{z}_0\vert^2}\Gamma_1\tilde{\psi}_1\right).
\end{equation}

From these preliminary calculations, we derive an expression for the superderivative of the projector $\mathbb{P}_1$:
\begin{equation}
\check{\partial}_+\mathbb{P}_1=\frac{1}{\vert \tilde{z}_1\vert^2}(\mathbb{I}-\mathbb{P}_0-\mathbb{P}_1)\left(\Gamma_2\tilde{\psi}_2-\frac{\tilde{z}_0^{\dagger}\tilde{\psi}_1}{\vert \tilde{z}_0\vert^2}\Gamma_1\tilde{\psi}_1\right)\tilde{z}_1^{\dagger}-\frac{\tilde{z}_1^{\dagger}(\Gamma_1\tilde{\psi}_1)}{\vert\tilde{z}_0\vert^2\vert\tilde{z}_1\vert^2}\tilde{z}_1 \tilde{z}_0^{\dagger}.
\end{equation}
From this, we now have an explicit expression for $\Xi_1^{\dagger}$ to be put into (\ref{susycons}):
\begin{equation}
\Xi_1^{\dagger}=\frac{1}{\vert \tilde{z}_1\vert^2}(\mathbb{I}-\mathbb{P}_0-\mathbb{P}_1)\left(\Gamma_2\tilde{\psi}_2-\frac{\tilde{z}_0^{\dagger}\tilde{\psi}_1}{\vert \tilde{z}_0\vert^2}\Gamma_1\tilde{\psi}_1\right)\tilde{z}_1^{\dagger}+\frac{\tilde{z}_1^{\dagger}(\Gamma_1\tilde{\psi}_1)}{\vert\tilde{z}_0\vert^2\vert\tilde{z}_1\vert^2}\tilde{z}_1 \tilde{z}_0^{\dagger}
\end{equation}
and we note that
\begin{equation}
\mathbb{P}_1\check{\partial}_+\mathbb{P}_0=\frac{\tilde{z}_1^{\dagger}(\Gamma_1\tilde{\psi}_1)}{\vert\tilde{z}_0\vert^2\vert\tilde{z}_1\vert^2}\tilde{z}_1 \tilde{z}_0^{\dagger}.\label{P1DP0}
\end{equation}
As a consequence we may rewrite $\Xi_1^{\dagger}$ as
\begin{equation}
\Xi_1^{\dagger}=\check{\partial}_+\mathbb{P}_1+2\mathbb{P}_1\check{\partial}_+\mathbb{P}_0.\label{noconstraintcons}
\end{equation}
From this, we see that the Euler-Lagrange equations (\ref{susycons}) reduce to having to satisfy:
\begin{equation}
\check{\partial}_+\Xi_1+\check{\partial}_-\Xi_1^{\dagger}=\check{\partial}_+((\check{\partial}_-\mathbb{P}_0)\mathbb{P}_1)+\check{\partial}_-(\mathbb{P}_1\check{\partial}_+\mathbb{P}_0)=0. \label{ELP1}
\end{equation}

We will show in Proposition 1 that this condition is equivalent to the constraint $\Gamma_1\tilde{\psi}_1\in\hbox{ker}(\mathbb{I}-\mathbb{P}_0-\mathbb{P}_1)$, \textit{i.e.} $(\mathbb{I}-\mathbb{P}_0-\mathbb{P}_1)(\Gamma_1\tilde{\psi}_1)=0$.

Before we do this let us note that for the projector $\mathbb{P}_2$, we can use a result proved in \cite{delisle}, that demonstrates that $\mathbb{P}_0+\mathbb{P}_1$ is a holomorphic solution of the supersymmetric $G(2,3)$ model and so, using the completeness relation (\ref{complete}), $\mathbb{P}_2$ also solves the Euler-Lagrange equations. Furthermore, the projector $\mathbb{P}_2$ corresponds to an anti-holomorphic solution in the sense that it satisfies $(\check{\partial}_+\mathbb{P}_2)\mathbb{P}_2=0$. Again, using the completeness relation (\ref{complete}), we get
\begin{equation}
(\check{\partial}_+\mathbb{P}_2)\mathbb{P}_2=-\check{\partial}_+(\mathbb{P}_0+\mathbb{P}_1)(\mathbb{I}-(\mathbb{P}_0+\mathbb{P}_1))=-\check{\partial}_+(\mathbb{P}_0+\mathbb{P}_1)+\check{\partial}_+(\mathbb{P}_0+\mathbb{P}_1)(\mathbb{P}_0+\mathbb{P}_1)
\end{equation}
and the result follows from the fact that $\mathbb{P}_0+\mathbb{P}_1$, being holomorphic, satisfies the identity $\check{\partial}_+(\mathbb{P}_0+\mathbb{P}_1)(\mathbb{P}_0+\mathbb{P}_1)=\check{\partial}_+(\mathbb{P}_0+\mathbb{P}_1)$.

\vskip 0.2cm

\noindent\textbf{Proposition 1:} The projector $\mathbb{P}_1$ solves  the Euler-Lagrange equations (\ref{EL}) if and only if 
\begin{equation}
\Gamma_1\tilde{\psi}_1\in\hbox{ker}(\mathbb{I}-\mathbb{P}_0-\mathbb{P}_1).\label{constraintCP2}
\end{equation}

\noindent\textbf{Proof:} 
By direct computation, we can show that (\ref{ELP1}) is equivalent to the following equation
\begin{equation}
\left(a_1+\alpha_1\frac{(\Gamma_1\tilde{\psi}_1)^{\dagger}\tilde{z}_0}{\vert \tilde{z}_0\vert^2}\right)\frac{\tilde{z}_1 \tilde{z}_0^{\dagger}}{\vert\tilde{z}_0\vert^2}-\left(a_1^{\dagger}+\frac{\tilde{z}_0^{\dagger}(\Gamma_1\tilde{\psi}_1)}{\vert\tilde{z}_0\vert^2}\alpha_1^{\dagger}\right)\frac{\tilde{z}_0 \tilde{z}_1^{\dagger}}{\vert\tilde{z}_0\vert^2}-\alpha_1\frac{\tilde{z}_1 (\Gamma_1\tilde{\psi}_1)^{\dagger}}{\vert\tilde{z}_0\vert^2}
-\alpha_1^{\dagger}\frac{(\Gamma_1\tilde{\psi}_1) \tilde{z}_1^{\dagger}}{\vert\tilde{z}_0\vert^2}=0,\label{noconstrainteq}
\end{equation}
where $a_1$ and $\alpha_1$ are, respectively, bosonic and fermionic functions defined by
\begin{equation}
a_1^{\dagger}=(\Gamma_1\tilde{\psi}_1)^{\dagger}\check{\partial}_+\left(\frac{\tilde{z}_1}{\vert\tilde{z}_1\vert^2}\right),\quad \alpha_1=\frac{\tilde{z}_1^{\dagger}(\Gamma_1\tilde{\psi}_1)}{\vert\tilde{z}_1\vert^2}.
\label{eee}
\end{equation}
We easily see from these expressions that $a_1=\check{\partial}_-\alpha_1$. 

We can now act on (\ref{noconstrainteq})  from the left with $\tilde{z}_0^{\dagger}$. It leads to (recall that $\tilde{z}_1^{\dagger}\tilde{z}_0=0$ and, that $\Gamma_1\tilde{\psi}_1$ and $\alpha_1^{\dagger}$ are fermionic and so anticommute)
\begin{equation}
a_1\tilde{z}_1=0\quad \Longrightarrow\quad a_1=0,
\end{equation}
which is equivalent to 
\begin{equation}
\left(\Gamma_2\tilde{\psi}_2-\frac{\tilde{z}_0^{\dagger}\tilde{\psi}_1}{\vert\tilde{z}_0\vert^2}\Gamma_1\tilde{\psi}_1\right)^\dagger (\mathbb{I}-\mathbb{P}_0-\mathbb{P}_1)(\Gamma_1\tilde{\psi}_1)=0. \label{a1equal0}
\end{equation}

So, putting $a_1=0$, (\ref{noconstrainteq}) becomes
\begin{equation}
\alpha_1 \tilde{z}_1(\Gamma_1\tilde{\psi}_1)^{\dagger}(\mathbb{P}_0-\mathbb{I})+\alpha_1^{\dagger}(\mathbb{P}_0-\mathbb{I})(\Gamma_1\tilde{\psi}_1) \tilde{z}_1^{\dagger}=0.
\end{equation}
Acting on this expression from the right with $\tilde{z}_1$ and using the fact that $\mathbb{P}_1(\Gamma_1\tilde{\psi}_1)=\alpha_1\tilde{z}_1$ we get
\begin{equation}
\alpha_1^{\dagger}(\mathbb{I}-\mathbb{P}_0-\mathbb{P}_1)(\Gamma_1\tilde{\psi}_1)=0.
\end{equation}
The equation is clearly satisfied if $\alpha_1=0$. However, this choice has to be rejected. Indeed, $\alpha_1=0$ is  equivalent to
\begin{equation}
\tilde{z}_1^{\dagger}(\Gamma_1\tilde{\psi}_1)=0
\end{equation}
and, from (\ref{derimoinsz1}), we would have then got  that $\check{\partial}_-\tilde{z}_1=0$. However, this  would have meant that the supervector $\tilde{z}_1=\tilde{z}_1(x_+,\theta_+)$ is holomorphic but we want $\tilde{z}_1$ to be non-holomorphic. Thus  we require $\alpha_1\neq 0$. This leads immediately to the constraint (\ref{constraintCP2}). Moreover, in this case the equation (\ref{a1equal0}) is automatically satisfied. Thus we have shown that in order for $\mathbb{P}_1$ to be a solution of the Euler-Lagrange equation, we require the constraint (\ref{constraintCP2}) to be satisfied.

Finally, we have to show that if the constraint (\ref{constraintCP2}) is satisfied  $\mathbb{P}_1$ is a solution of the Euler-Lagrange equation. Indeed,
 the constraint (\ref{constraintCP2}) can be rewritten as
 \begin{equation}
\mathbb{P}_1(\Gamma_1\tilde{\psi}_1)=(\mathbb{I}-\mathbb{P}_0)(\Gamma_1\tilde{\psi}_1)
\end{equation}
from which we deduce, from the expression (\ref{P1DP0}), that
\begin{equation}
\mathbb{P}_1\check{\partial}_+\mathbb{P}_0=\mathbb{P}_1(\Gamma_1\tilde{\psi}_1)\frac{\tilde{z}_0^{\dagger}}{\vert\tilde{z}_0\vert^2}=(\mathbb{I}-\mathbb{P}_0)(\Gamma_1\tilde{\psi}_1)\frac{\tilde{z}_0^{\dagger}}{\vert\tilde{z}_0\vert^2}=\check{\partial}_+\mathbb{P}_0.
\end{equation}
From this last result, we see that $\Xi_1^{\dagger}=\check{\partial}_+(\mathbb{P}_1+2\mathbb{P}_0)$ and so, using the fact that $\{\check{\partial}_+,\check{\partial}_-\}=0$, we conclude that $\mathbb{P}_1$ is indeed a solution of the Euler-Lagrange equation.

This concludes the proof.

\subsection{The study of the constraint}

Here we look at the constraint (\ref{constraintCP2}). 

The constraint (\ref{constraintCP2}) can be equivalently written as
\begin{equation}
(\mathbb{P}_0+\mathbb{P}_1)(\Gamma_1\tilde{\psi}_1)=\Gamma_1\tilde{\psi}_1,
\end{equation}
which implies that $\Gamma_1\tilde{\psi}_1$ is an eigenvector of the matrix $\mathbb{P}_0+\mathbb{P}_1$ of eigenvalue 1. The next step is to show that the matrix $\mathbb{P}_0+\mathbb{P}_1$ is diagonalisable. We note that 
\begin{equation}
(\mathbb{P}_0+\mathbb{P}_1)\tilde{\psi}_j=\tilde{\psi}_j,\quad (\mathbb{P}_0+\mathbb{P}_1)\tilde{z}_2=0
\end{equation}
for $j=0,1$ and, by construction, the vectors $\tilde{\psi}_0$, $\tilde{\psi}_1$ and $\tilde{z}_2$ are linearly independent implying that the matrix is diagonalisable. From the constraint (\ref{constraintCP2}), we deduce that $\Gamma_1\tilde{\psi}_1$ lies in the eigenspace of $(\mathbb{P}_0+\mathbb{P}_1)$ and its eigenvalue is 1. So we can write
\begin{equation}
\Gamma_1\tilde{\psi}_1=\alpha_0\tilde{\psi}_0+\alpha_1\tilde{\psi}_1,
\end{equation}
where $\alpha_0$ and $\alpha_1$ are fermionic functions of $(x_+,\theta_+)$.

For $\Gamma_2\tilde{\psi}_2$ we can use the completeness relation, {\it i.e.} that vectors 
$\tilde{\psi}_i$, $i=1,2,3$ can be taken as basis vectors of $\mathbb{C}^3$, to expand
\begin{equation}
\Gamma_2\tilde{\psi}_2=\beta_0\tilde{\psi}_0+\beta_1\tilde{\psi}_1+\beta_2\tilde{\psi}_2,
\end{equation}
where $\beta_0$, $\beta_1$ and $\beta_2$ are fermionic functions of $(x_+,\theta_+)$. So, finding the solutions of the system (\ref{systemCP2}) together with the constraint (\ref{constraintCP2}) is equivalent to finding the solutions of the system
\begin{equation}
\alpha_0\tilde{\psi}_0+\alpha_1\tilde{\psi}_1=\check{\partial}_+\tilde{\psi}_0,\quad \beta_0\tilde{\psi}_0+\beta_1\tilde{\psi}_1+\beta_2\tilde{\psi}_2=\check{\partial}_+\tilde{\psi}_1.\label{newsystemCP2}
\end{equation}
This we do in the next section.

\subsection{General construction of $\tilde{\psi}_0$, $\tilde{\psi}_1$ and $\tilde{\psi}_2$}

Here we determine the general expressions for the superfields $\tilde{\psi}_0$, $\tilde{\psi}_1$ and $\tilde{\psi}_2$ of the system (\ref{newsystemCP2}). To do so, we put
\begin{equation}
\alpha_j=\alpha_j^f+i\theta_+\alpha_j^b,\quad \beta_k=\beta_k^f+i\theta_+\beta_k^b,\quad \tilde{\psi}_k=\tilde{\psi}_k^b+i\theta_+\tilde{\psi}_k^f
\end{equation}
for $j=0,1$ and $k=0,1,2$. The superscripts $f$ and $b$ refers to, respectively, to fermionic and bosonic quantities. From these expressions, we may rewrite system (\ref{newsystemCP2}) in component form as a set of four equations:
\begin{eqnarray}
\alpha_0^f\tilde{\psi}_0^b+\alpha_1^f\tilde{\psi}_1^b&=&\tilde{\psi}_0^f,\label{eq1}\\
\alpha_0^b\tilde{\psi}_0^b+\alpha_1^b\tilde{\psi}_1^b-\alpha_0^f\tilde{\psi}_0^f-\alpha_1^f\tilde{\psi}_1^f&=&-i\partial_+\tilde{\psi}_0^b,\label{eq2}\\
\beta_0^f\tilde{\psi}_0^b+\beta_1^f\tilde{\psi}_1^b+\beta_2^f\tilde{\psi}_2^b&=&\tilde{\psi}_1^f,\label{eq3}\\
\beta_0^b\tilde{\psi}_0^b+\beta_1^b\tilde{\psi}_1^b+\beta_2^b\tilde{\psi}_2^b-\beta_0^f\tilde{\psi}_0^f-\beta_1^f\tilde{\psi}_1^f-\beta_2^f\tilde{\psi}_2^f&=&-i\partial_+\tilde{\psi}_1^b.\label{eq4}
\end{eqnarray}
The details of the above resolution are given in Appendix A and, as it is shown there, the general solution of this system may be expressed in terms of the field $\tilde{\psi}_0^b$ and its consecutive ordinary derivatives. Indeed, the expressions for $\tilde{\psi}_0^f$ and $\tilde{\psi}_1^b$ are given by
\begin{eqnarray}
\tilde{\psi}_0^f&=&\left(\alpha_0^f-\frac{\alpha_1^f\alpha_0^b}{\alpha_1^b}\right)\tilde{\psi}_0^b-i\frac{\alpha_1^f}{\alpha_1^b}\partial_+\tilde{\psi}_0^b,\\
\tilde{\psi}_1^b&=&A_0\tilde{\psi}_0^b+A_1\partial_+\tilde{\psi}_0^b-i\frac{\alpha_1^f\beta_2^f}{\alpha_1^b\beta_2^b}\left((\partial_+A_0)\tilde{\psi}_0^b+(A_0+\partial_+A_1)\partial_+\tilde{\psi}_0^b+A_1\partial_+^2\tilde{\psi}_0^b\right),\label{psi1bnew}
\end{eqnarray}
where the quantities $A_0$ and $A_1$, in their explicit form, are given by
\begin{eqnarray}
A_0&=&-\frac{\alpha_0^b}{\alpha_1^b}\left(1+\frac{\alpha_0^f\alpha_1^f}{\alpha_1^b}\right)+\frac{\alpha_1^f}{\alpha_1^b}\left(\beta_0^f-\frac{\beta_2^f\beta_0^b}{\beta_2^b}+\frac{\beta_2^f\beta_1^f\beta_0^f}{\beta_2^b}\right)-\frac{\alpha_0^b\alpha_1^f}{(\alpha_1^b)^2}\left(\beta_1^f-\frac{\beta_2^f\beta_1^b}{\beta_2^b}\right)\nonumber\\&+&\frac{\alpha_1^f\beta_2^f\beta_0^f}{\alpha_1^b\beta_2^b}\left(\alpha_0^f-\frac{\alpha_1^f\alpha_0^b}{\alpha_1^b}\right)\label{A0},\\
A_1&=&-\frac{i}{\alpha_1^b}\left(1+\frac{\alpha_0^f\alpha_1^f}{\alpha_1^b}+\frac{\alpha_1^f}{\alpha_1^b}\left(\beta_1^f-\frac{\beta_2^f\beta_1^b}{\beta_2^b}\right)\right).\label{A1}
\end{eqnarray}
Moreover, the explicit expression of the field $\tilde{\psi}_1^f$ can be given in terms of the fields $\tilde{\psi}_0^b$, $\tilde{\psi}_0^f$ and $\tilde{\psi}_1^b$ and it takes the form
\begin{equation}
\tilde{\psi}_1^f=\left(\beta_0^f-\frac{\beta_2^f\beta_0^b}{\beta_2^b}+\frac{\beta_2^f\beta_1^f\beta_0^f}{\beta_2^b}\right)\tilde{\psi}_0^b+\left(\beta_1^f-\frac{\beta_2^f\beta_1^b}{\beta_2^b}\right)\tilde{\psi}_1^b+\frac{\beta_2^f\beta_0^f}{\beta_2^b}\tilde{\psi}_0^f-i\frac{\beta_2^f}{\beta_2^b}\partial_+\tilde{\psi}_1^b.
\end{equation}
The above expressions can then be used to determine the field $\tilde{\psi}_2^b$ and $\tilde{\psi}_2^f$ although we do not really need them as, in fact, we are interested in the form of $\Phi_2$ which provides the last ({\it i.e.} antiholomorphic) solution of the model.
$\Phi_2$ is, however, orthogonal to $\Phi_1$
and so can be calculated by taking, as an example, $\tilde{\psi}_2=\tilde{\psi}_1\times \tilde{\psi}_0$ where $\times$ denotes the vector product. We like to point out that this particular choice is restrictive in the sense that $\tilde{\psi}_2^f$ is completely determined in opposition with the general case where $\tilde{\psi}_2^f$ is a free field. 
Note that we stated just before (just before our proposition 1) that $\Phi_2$ is an antiholomorphic solution.

The details of all these calculations are presented in Appendix A. In the next section, we explore some special cases of the obtained solutions.


\subsection{Special solutions and connection with the operator $P_{x_+}$}

In this section, we discuss, in more detail, a special case of the solution obtained in the previous section, in which we have made the particular choice $\alpha_0=\beta_0=\beta_1=0$,  \textit{i.e.}
\begin{equation}
\Gamma_1\tilde{\psi_1}=\alpha_1\tilde{\psi}_1,\quad \Gamma_2\tilde{\psi}_2=\beta_2\tilde{\psi}_2.\label{specialsys}
\end{equation}
In this case, the quantities $A_0$ and $A_1$ take the simple forms
\begin{equation}
A_0=0,\quad A_1=-\frac{i}{\alpha_1^b}
\end{equation}
and, in consequence, the fields $\tilde{\psi}_0^f$, $\tilde{\psi}_1^b$, $\tilde{\psi}_1^f$ and $\tilde{\psi}_2^b$ are given by
\begin{equation}
\tilde{\psi}_0^f=-i\frac{\alpha_1^f}{\alpha_1^b}\partial_+\tilde{\psi}_0^b,\quad \tilde{\psi}_1^f=-i\frac{\beta_2^f}{\beta_2^b}\partial_+\tilde{\psi}_1^b,\quad \tilde{\psi}_2^b=\frac{\beta_2^f}{\beta_2^b}\tilde{\psi}_2^f-\frac{i}{\beta_2^b}\partial_+\tilde{\psi}_1^b,
\end{equation}
together with
\begin{equation}
\tilde{\psi}_1^b=\left(-\frac{i}{\alpha_1^b}+\frac{\alpha_1^f\beta_2^f\partial_+\alpha_1^b}{(\alpha_1^b)^3\beta_2^b}\right)\partial_+\tilde{\psi}_0^b-\frac{\alpha_1^f\beta_2^f}{(\alpha_1^b)^2\beta_2^b}\partial_+^2\tilde{\psi}_0^b.
\end{equation}

\noindent\textbf{Remark:} One important thing to stress here is that the obtained solutions in this paper are more general then the one discussed in the \cite{delisle}. One way of seeing this involves looking at the expression for $\tilde{z}_1$, for the system (\ref{specialsys}), in which we set $\theta_+=\theta_-=0$. For $\tilde{z}_0=\tilde{\psi}_0$, let us use the following notation
\begin{equation}
\tilde{z}_0\vert_{\theta_+=\theta_-=0}=\tilde{\psi}_0^{b}=u(x_+).
\end{equation}
In this case we find that the bosonic part of $\tilde{z}_1$ is given by
\begin{equation}
\tilde{z}_1\vert_{\theta_+=\theta_-=0}=\left(-\frac{i}{\alpha_1^{b}}+\frac{\alpha_1^{f}\beta_2^{f}}{(\alpha_1^{b})^3\beta_2^{b}}\partial_+\alpha_1^{b}\right)P_{x_+}u-\frac{\alpha_1^{f}\beta_2^{f}}{(\alpha_1^{b})^2\beta_2^{b}}\left(\mathbb{I}-\frac{uu^{\dagger}}{\vert u\vert^2}\right)\partial_+^2u.
\end{equation}
This expression is different then the one we have obtained in \cite{delisle}. Indeed, in \cite{delisle} we had $\tilde{z}_1\vert_{\theta_+=\theta_-=0}=-\frac{i}{\alpha_1^{b}}P_{x_+}u$, where $P_{x_+}u=\partial_+u-\frac{u^{\dagger}\partial_+u}{\vert u\vert^2}u$.

As a final comment, it might be worthwhile to give the explicit expressions of $\tilde{\psi}_0$, $\tilde{\psi}_1$ and $\tilde{\psi}_2$ in terms of the free components $\tilde{\psi}_0^{b}$ and $\tilde{\psi}_2^{f}$. We have
\begin{eqnarray}
\tilde{\psi}_0&=&\tilde{\psi}_0^{b}+\theta_+\frac{\alpha_1^{f}}{\alpha_1^{b}}\partial_+\tilde{\psi}_0^{b},\\
\tilde{\psi}_1&=&-\frac{i}{\alpha_1^{b}}\partial_+\tilde{\psi}_0^{b}+\frac{\beta_2^{f}}{\beta_2^{b}}\left(\frac{\alpha_1^{f}}{\alpha_1^{b}}+i\theta_+\left(1-i\frac{\alpha_1^{f}\partial_+\beta_2^{f}}{\alpha_1^{b}\beta_2^{b}}\right)\right)\partial_+\left(\frac{1}{\alpha_1^{b}}\partial_+\tilde{\psi}_0^{b}\right),\\
\tilde{\psi}_2&=&-\frac{i}{\beta_2^{b}}\partial_+\left(-\frac{i}{\alpha_1^{b}}\partial_+\tilde{\psi}_0^{b}-\frac{\alpha_1^{f}\beta_2^{f}}{\alpha_1^{b}\beta_2^{b}}\partial_+\left(\frac{1}{\alpha_1^{b}}\partial_+\tilde{\psi}_0^{b}\right)\right)+\frac{\beta_2^{f}+i\theta_+\beta_2^{b}}{\beta_2^{b}}\tilde{\psi}_2^{f}.
\end{eqnarray}

\section{The general model}
In this section, we generalise the results of the $\mathbb{C}P^2$ model to the general $\mathbb{C}P^{N-1}$ model. In order to construct these solutions, we consider a set of $N$ bosonic holomorphic supervectors of $\mathbb{C}^{N}$ given by $\{\tilde{\psi}_0,\tilde{\psi}_1,\cdots,\tilde{\psi}_{N-1}\}$ which are a solution of the system of equations
\begin{equation}
\tilde{\psi}_0,\quad \Gamma_1\tilde{\psi}_1=\check{\partial}_+\tilde{\psi}_0,\quad \Gamma_2\tilde{\psi}_2=\check{\partial}_+\tilde{\psi}_1,\cdots,\quad\Gamma_{N-1}\tilde{\psi}_{N-1}=\check{\partial}_+\tilde{\psi}_{N-2},\label{holomorphicsys}
\end{equation}
where $\Gamma_j$, for $j=1,2,\cdots,N-1$, are $N\times N$ fermionic valued matrices \cite{cornwell} of $(x_+,\theta_+)$. Then, we perform  the Gram-Schmidt orthogonalisation procedure of this set of supervectors and obtain a new set given by
\begin{equation}
\tilde{z}_0=\tilde{\psi}_0,\quad \tilde{z}_j=\left(\mathbb{I}-\sum_{k=0}^{j-1}\mathbb{P}_k\right)\tilde{\psi}_j,\quad \mathbb{P}_j=\frac{\tilde{z}_j \tilde{z}_j^{\dagger}}{\vert\tilde{z}_j\vert^2}\label{solutionCPN}
\end{equation}
for $j=1,2,\cdots, N-1$.

 We now have the following proposition:
 
\noindent\textbf{Proposition 3:} If $\Gamma_j\psi_j$ of the original set 
(\ref{holomorphicsys})
satisfy the constraints
\begin{equation}
\Gamma_j\tilde{\psi}_j\in\ker\left(\mathbb{I}-\sum_{k=0}^{j}\mathbb{P}_k\right),\quad j=1,2,\cdots,N-2,\label{generalconstraint}
\end{equation}
then the supervectors $\{\tilde{z}_0,\tilde{z}_1,\cdots,\tilde{z}_{N-1}\}$ possess the following properties
\begin{equation}
\check{\partial}_-\tilde{z}_j=-\frac{(\Gamma_j\tilde{\psi}_j)^{\dagger}\tilde{z}_j}{\vert \tilde{z}_{j-1}\vert^2}\tilde{z}_{j-1},\quad \check{\partial}_+\left(\frac{\tilde{z}_j}{\vert\tilde{z}_j\vert^2}\right)=\frac{1}{\vert\tilde{z}_j\vert^2}\left(\mathbb{I}-\sum_{k=0}^j\mathbb{P}_k\right)\Gamma_{j+1}\tilde{\psi}_{j+1}\label{generalproperties}
\end{equation}
for $j=0,1,\cdots, N-1$.

\noindent\textbf{Proof:} To prove these properties we proceed by induction. The first step of the proof were performed in section 3.1 and, thus, we can go forward to the induction part of the proof. Let us assume that the properties hold for $0\leq j\leq m$. Before moving on, let us perform some additional preliminary calculations. We have
\begin{equation}
\check{\partial}_+\mathbb{P}_j=\frac{1}{\vert\tilde{z}_j\vert^2}\left(\mathbb{I}-\sum_{k=0}^{j}\mathbb{P}_k\right)(\Gamma_{j+1}\tilde{\psi}_{j+1}) \tilde{z}_j^{\dagger}-\frac{\tilde{z}_j^{\dagger}(\Gamma_j\tilde{\psi}_j)}{\vert\tilde{z}_{j-1}\vert^2\vert\tilde{z}_j\vert^2}\tilde{z}_j\tilde{z}_{j-1}^{\dagger}.
\end{equation}
next we re-express the last term of this expression as
\begin{equation}
\frac{\tilde{z}_j^{\dagger}(\Gamma_j\tilde{\psi}_j)}{\vert\tilde{z}_{j-1}\vert^2\vert\tilde{z}_j\vert^2}\tilde{z}_j\tilde{z}_{j-1}^{\dagger}=\mathbb{P}_j(\Gamma_j\tilde{\psi}_j)\frac{\tilde{z}_{j-1}^{\dagger}}{\vert\tilde{z}_{j-1}\vert^2}.
\end{equation}
Using the constraint (\ref{generalconstraint}), we find that
\begin{equation}
\mathbb{P}_j(\Gamma_j\tilde{\psi}_j)=\left(\mathbb{I}-\sum_{k=0}^{j-1}\mathbb{P}_k\right)\Gamma_j\tilde{\psi}_j
\end{equation}
and we find that
\begin{equation}
\check{\partial}_+\mathbb{P}_j=\mathbb{B}_j-\mathbb{B}_{j-1},\quad \mathbb{B}_m=\frac{1}{\vert \tilde{z}_m\vert^2}\left(\mathbb{I}-\sum_{k=0}^m\mathbb{P}_k\right)(\Gamma_{m+1}\tilde{\psi}_{m+1}) \tilde{z}_m^{\dagger}.
\end{equation}
This allows us to observe that
\begin{equation}
\sum_{k=0}^{j}\check{\partial}_+\mathbb{P}_k=\sum_{k=0}^j(\mathbb{B}_k-\mathbb{B}_{k-1})=\mathbb{B}_j.
\end{equation}
We are now ready to proceed with the induction process and, using the above expression, we find that
\begin{equation}
\check{\partial}_-\tilde{z}_{m+1}=-\left(\sum_{j=0}^m\check{\partial}_-\mathbb{P}_j\right)\tilde{\psi}_{m+1}=-\frac{1}{\vert\tilde{z}_m\vert^2}\tilde{z}_m (\Gamma_{m+1}\tilde{\psi}_{m+1})^{\dagger}\left(\mathbb{I}-\sum_{j=0}^m\mathbb{P}_j\right)\tilde{\psi}_{m+1}.
\end{equation}

From the definition of $\tilde{z}_{m+1}$, we  get 
\begin{equation}
\check{\partial}_-\tilde{z}_{m+1}=-\frac{(\Gamma_{m+1}\tilde{\psi}_{m+1})^{\dagger}\tilde{z}_{m+1}}{\vert\tilde{z}_m\vert^2}\tilde{z}_m.
\end{equation}

This last result gives us that $\tilde{z}_{m+1}^{\dagger}\check{\partial}_-\tilde{z}_{m+1}=0$ from the orthogonality of $\tilde{z}_m$ and $\tilde{z}_{m+1}$ which in turn allows us to write
\begin{equation}
\check{\partial}_+\left(\frac{\tilde{z}_{m+1}}{\vert\tilde{z}_{m+1}\vert^2}\right)=\frac{1}{\vert\tilde{z}_{m+1}\vert^2}(\mathbb{I}-\mathbb{P}_{m+1})\check{\partial}_+\tilde{z}_{m+1}.
\end{equation}

Next we calculate the superderivative of $\tilde{z}_{m+1}$ and get
\begin{equation}
\check{\partial}_+\tilde{z}_{m+1}=-\left(\sum_{j=0}^m\check{\partial}_+\mathbb{P}_j\right)\tilde{\psi}_{m+1}+\left(\mathbb{I}-\sum_{j=0}^{m}\mathbb{P}_j\right)\Gamma_{m+2}\tilde{\psi}_{m+2}.
\end{equation}
Using these results we get a new expression
\begin{equation}
\check{\partial}_+\tilde{z}_{m+1}=\left(\mathbb{I}-\sum_{j=0}^{m}\mathbb{P}_j\right)\left(\Gamma_{m+2}\tilde{\psi}_{m+2}-\frac{\tilde{z}_m^{\dagger}\tilde{\psi}_{m+1}}{\vert\tilde{z}_m\vert^2}\Gamma_{m+1}\tilde{\psi}_{m+1}\right)
\end{equation}
and then, using the fact that $(\mathbb{I}-\mathbb{P}_{m+1})(\mathbb{I}-\sum_{j=0}^m\mathbb{P}_j)=\mathbb{I}-\sum_{j=0}^{m+1}\mathbb{P}_j$, we find that
\begin{equation}
\check{\partial}_+\left(\frac{\tilde{z}_{m+1}}{\vert\tilde{z}_{m+1}\vert^2}\right)=\frac{1}{\vert\tilde{z}_{m+1}\vert^2}\left(\mathbb{I}-\sum_{j=0}^{m+1}\mathbb{P}_j\right)\left(\Gamma_{m+2}\tilde{\psi}_{m+2}-\frac{\tilde{z}_m^{\dagger}\tilde{\psi}_{m+1}}{\vert\tilde{z}_m\vert^2}\Gamma_{m+1}\tilde{\psi}_{m+1}\right).
\end{equation}

The final result follows from constraint (\ref{generalconstraint}),
\begin{equation}
\check{\partial}_+\left(\frac{\tilde{z}_{m+1}}{\vert\tilde{z}_{m+1}\vert^2}\right)=\frac{1}{\vert\tilde{z}_{m+1}\vert^2}\left(\mathbb{I}-\sum_{j=0}^{m+1}\mathbb{P}_j\right)\left(\Gamma_{m+2}\tilde{\psi}_{m+2}\right).
\end{equation}
This concludes the proof.

We can now also prove the following theorem:

\noindent\textbf{Theorem:} If the constraints (\ref{generalconstraint}) are satisfied, the rank one orthogonal projectors $\mathbb{P}_j$ defined as in (\ref{solutionCPN}) solve the Euler-Lagrange equations for $j=0,1,\cdots, N-1$.

\noindent\textbf{Proof:} Again, it is easy to show that
\begin{equation}
\Xi_{j}^{\dagger}=[\check{\partial}_+\mathbb{P}_j,\mathbb{P}_j]=\mathbb{B}_j+\mathbb{B}_{j-1}
\end{equation}
and, in order to prove the theorem, we have re-express the above expression. We note that
\begin{equation}
\mathbb{B}_j+\mathbb{B}_{j-1}=(\mathbb{B}_j-\mathbb{B}_{j-1})+2(\mathbb{B}_{j-1}-\mathbb{B}_{j-2})+\cdots+2(\mathbb{B}_1-\mathbb{B}_0)+2\mathbb{B}_0
\end{equation}
and this leads to
\begin{equation}
\Xi_j^{\dagger}=\check{\partial}_+\left(\mathbb{P}_j+2\sum_{k=0}^{j-1}\mathbb{P}_k\right).
\end{equation}
This concludes our proof.

\textbf{Example:} We can solve the holomorphic system (\ref{holomorphicsys}) in the particular case when the constraints (\ref{generalconstraint}) are trivially satisfied. As an example, we choose $\Gamma_j(x_+,\theta_+)=\eta(x_+,\theta_+)\mathbb{I}$ where $\eta(x_+,\theta_+)=\eta^{(f)}(x_+)+i\theta_+\eta^{(b)}$. For simplicity, we have chosen the bosonic part of $\eta$ as a constant and the fermionic part as an odd function of $x_+$. In this particular case, the system (\ref{holomorphicsys}) possesses the general solution
\begin{eqnarray}
\tilde{\psi}_j^{(b)}&=&\left(-\frac{i}{\eta^{(b)}}\right)^j\partial_+^j\tilde{\psi}_0^{(b)},\quad j=1,2,\cdots,N-2,\\
\tilde{\psi}_{N-1}^{(b)}&=&\frac{\eta^{(f)}}{\eta^{(b)}}\tilde{\psi}_{N-1}^{(f)}+\left(-\frac{i}{\eta^{(b)}}\right)^{N-1}\partial_+^{N-1}\tilde{\psi}_0^{(b)},\\
\tilde{\psi}_j^{(f)}&=&\eta^{(f)}\left(-\frac{i}{\eta^{(b)}}\right)^{j+1}\partial_+^{j+1}\tilde{\psi}_0^{(b)},\quad j=0,1,\cdots, N-2.
\end{eqnarray}
These solutions are exactly the ones considered in \cite{delisle} and thus we see again that our procedure reproduces and generalises the previously found general solutions.

\subsection{The constraints and system of equations}
Let us look deeper into the imposed constraints (\ref{generalconstraint}) and see what would be the conditions on the $\Gamma$ matrices in order that they are satisfied. 

The first observation that one can make is that the set of vectors $\{\tilde{\psi}_0,\tilde{\psi}_1,\cdots,\tilde{\psi}_{N-1}\}$ should be linearly independent in order for the vectors $\tilde{z}_j$ to be different from zero for all $j$. This fact can be easily seen by the construction of the $\tilde{z}_j$.

Let us re-write the general constraint (\ref{generalconstraint}) as
\begin{equation}
\left(\mathbb{I}-\sum_{k=0}^{j}\mathbb{P}_k\right)(\Gamma_j\tilde{\psi}_j)=0\quad \Longleftrightarrow\quad \left(\sum_{k=0}^j\mathbb{P}_k\right)(\Gamma_j\tilde{\psi}_j)=\Gamma_j\tilde{\psi}_j,
\end{equation}
which means that $\Gamma_j\tilde{\psi}_j$ is an eigenvector of eigenvalue 1 for the matrix $\sum_{k=0}^{j}\mathbb{P}_k$. Furthermore, we can make the observations 
\begin{equation}
\left(\sum_{k=0}^j\mathbb{P}_k\right)\tilde{\psi}_m=\tilde{\psi}_m,\quad \left(\sum_{k=0}^j\mathbb{P}_k\right)\tilde{z}_l=0
\end{equation}
for $0\leq m\leq j$ and $j<l\leq N-1$. Using the fact that all these vectors are linearly independent, we have that the matrix $\sum_{k=0}^j\mathbb{P}_k$ is diagonalizable and has two eigenspaces: the first one is spanned by $\{\tilde{\psi}_0,\tilde{\psi}_1,\cdots,\tilde{\psi}_j\}$ corresponding to eigenvalue 1 and the second is spanned by $\{\tilde{z}_{j+1},\tilde{z}_{j+2},\cdots,\tilde{z}_{N-1}\}$ for eigenvalue 0. These facts are true for all $j=0,1,\cdots,N-1$.

We may thus write
\begin{equation}
\Gamma_j\tilde{\psi}_j=\sum_{k=0}^j\alpha_k^{(j)}\tilde{\psi}_k,
\end{equation}
for $\alpha_k^{(j)}$ odd functions of $(x_+,\theta_+)$ and $j=1,2,\cdots,N-2$. Then, the general properties (\ref{generalproperties}) take nice closed forms as
\begin{equation}
\check{\partial}_-\tilde{z}_j=-(\alpha_j^{(j)})^{\dagger}\frac{\vert\tilde{z}_j\vert^2}{\vert\tilde{z}_{j-1}\vert^2}\tilde{z}_{j-1},\quad \check{\partial}_+\left(\frac{\tilde{z}_j}{\vert\tilde{z}_j\vert^2}\right)=\alpha_{j+1}^{(j+1)}\frac{\tilde{z}_{j+1}}{\vert\tilde{z}_j\vert^2},
\end{equation}
which, modulo the odd functions $\alpha$, are identical to the one obtained in the bosonic $\mathbb{C}P^{N-1}$ sigma model [1]. In this case, obtaining the solutions of the system of equation (\ref{holomorphicsys}) is equivalent to solve the new system
\begin{equation}
\sum_{k=0}^j\alpha_k^{(j)}\tilde{\psi}_k=\check{\partial}_+\tilde{\psi}_{j-1}
\end{equation}
for $j=1,2,\cdots,N-1$.

\section{Conclusion and future outlook}

In this paper, we have presented a new and systematic procedure for constructing holomorphic, non-holomorphic and anti-holomorphic solutions of the supersymmetric $\mathbb{C}P^{N-1}$ sigma models. Our procedure generalises the one used to construct the complete set of solutions of the bosonic model called the holomorphic method. Although we think we have obtained the complete set of solutions in the supersymmetric model, the proof of completeness is still missing and this is part of our future projects. 

To describe our construction, we have focused our attention on the $\mathbb{C}P^2$ supersymmetric model which is the simplest model which possesses all three kinds of solutions namely holomorphic, non-holomorphic and anti-holomorphic. The main point of our approach involved rewriting the system in the form (\ref{systemCP2}) and then rewriting the constraint in the form (\ref{constraintCP2}). The key observation here involved the realisation that the constraint can be interpreted in terms of eigenvectors of the matrix $\mathbb{P}_0+\mathbb{P}_1$. This observation has led to the complete decoupling of the system of  equations (\ref{systemCP2}) and so allowed us to find its  special solutions. The complete solution of the resultant set of equations  is straightforward but requires tedious calculations. These complications have prevented us from presenting general solutions of the system (\ref{holomorphicsys}) corresponding to the $\mathbb{C}P^{N-1}$ model for $N>3$. Fortunately, the constraints (\ref{generalconstraint}) have also found their interpretation in terms of eigenvectors and we believe that this key idea will lead to the more complete understanding of the $\mathbb{C}P^{N-1}$ system of equations. This is planned for our future work.

Another interesting fact about these new solutions is that they exhibit a real mixing of bosonic and fermionic degrees of freedom. This mixing is apparent in the discussed solutions in the $\mathbb{C}P^2$ model where products of fermionic quantities 
appear in the bosonic fields. This property was not present in our previous paper \cite{delisle} and we have described this difference in section 3.4.

In our future work, we also plan to investigate the surfaces that can be constructed from these new solutions and study their geometries. Such surfaces were extensively studied in the past for the bosonic $\mathbb{C}P^{N-1}$ model and for more general Grassmannians \cite{delisle1,delisle2,delisle3,grundland1,grundland,hussin,bolton,jiao,jiao1,jie,peng,peng1,hussin1}. In particular, in our previous papers \cite{delisle1,delisle2,delisle3}, we attempted to classify the solutions of general Grassmannian sigma models that correspond to surfaces of constant gaussian curvature. In the particular case of $\mathbb{C}P^{N-1}$, a complete classification was given in terms of the Veronese curve \cite{bolton}. In the supersymmetric case, we have a similar classification but only for the holomorphic case \cite{delisle,hussin1}. The challenge here resides in having to define the surfaces which correspond to the bosonic model but which are elements of the $su(N)$ Lie algebra.

 \section*{Acknowledgements}
LD acknowledges a Natural Sciences and Engineering Research Council of Canada (NSERC) postdoctoral fellowship. VH has been supported by research grants from NSERC.

\section*{Appendix A}

In this Appendix, we present some details of the calculations leading to the solutions of the system of equations (\ref{eq1}), (\ref{eq2}), (\ref{eq3}) and (\ref{eq4}). The first step involves finding a solution of equation (\ref{eq2}) with respect to the bosonic field $\tilde{\psi}_1^b$. We have
\begin{equation}
\tilde{\psi}_1^b=\frac{1}{\alpha_1^b}(\alpha_0^f\tilde{\psi}_0^f+\alpha_1^f\tilde{\psi}_1^f-\alpha_0^b\tilde{\psi}_0^b-i\partial_+\tilde{\psi}_0^b).
\end{equation}
Next, we go further and we introduce the above expression into  (\ref{eq1}). We get
\begin{equation}
\tilde{\psi}_0^f=\left(\alpha_0^f-\frac{\alpha_1^f\alpha_0^b}{\alpha_1^b}\right)\tilde{\psi}_0^b-i\frac{\alpha_1^f}{\alpha_1^b}\partial_+\tilde{\psi}_0^b+\frac{\alpha_1^f\alpha_0^f}{\alpha_1^b}\tilde{\psi}_0^f,
\end{equation}
where we have used the fact that $(\alpha_1^f)^2=0$. This equation is equivalent to
\begin{equation}
\left(1-\frac{\alpha_1^f\alpha_0^f}{\alpha_1^b}\right)\tilde{\psi}_0^f=\left(\alpha_0^f-\frac{\alpha_1^f\alpha_0^b}{\alpha_1^b}\right)\tilde{\psi}_0^b-i\frac{\alpha_1^f}{\alpha_1^b}\partial_+\tilde{\psi}_0^b
\label{aaa}
\end{equation}
and, we obtain an explicit expression for $\tilde{\psi}_0^f$ by multiplying (\ref{aaa}) by the conjugate of the coefficient of $\tilde{\psi}_0^f$ in (\ref{aaa}). We have
\begin{equation}
\tilde{\psi}_0^f=\left(1+\frac{\alpha_0^f\alpha_1^f}{\alpha_1^b}\right)\left(1-\frac{\alpha_1^f\alpha_0^f}{\alpha_1^b}\right)\tilde{\psi}_0^f=\left(\alpha_0^f-\frac{\alpha_1^f\alpha_0^b}{\alpha_1^b}\right)\tilde{\psi}_0^b-i\frac{\alpha_1^f}{\alpha_1^b}\partial_+\tilde{\psi}_0^b.
\end{equation}
Next we put the expression for $\tilde{\psi}_0^f$ into $\tilde{\psi}_1^b$ and we get
\begin{equation}
\tilde{\psi}_1^b=-\frac{1}{\alpha_1^b}\left(1+\frac{\alpha_0^f\alpha_1^f}{\alpha_1^b}\right)(\alpha_0^b\tilde{\psi}_0^b+i\partial_+\tilde{\psi}_0^b)+\frac{\alpha_1^f}{\alpha_1^b}\tilde{\psi}_1^f.\label{psi1b}
\end{equation}
This expression depends on $\tilde{\psi}_1^f$, which should be determined by equations (\ref{eq3}) and (\ref{eq4}).
To do this we follow the steps we used before for the first two equations and so we determine $\tilde{\psi}_2^b$ with equation (\ref{eq4}). We find
\begin{equation}
\tilde{\psi}_2^b=\frac{1}{\beta_2^b}(\beta_0^f\tilde{\psi}_0^f+\beta_1^f\tilde{\psi}_1^f+\beta_2^f\tilde{\psi}_2^f-\beta_0^b\tilde{\psi}_0^b-\beta_1^b\tilde{\psi}_1^b-i\partial_+\tilde{\psi}_1^b),
\end{equation}
which we then insert into equation (\ref{eq3}) to get
\begin{equation}
\tilde{\psi}_1^f=\left(\beta_0^f-\frac{\beta_2^f\beta_0^b}{\beta_2^b}\right)\tilde{\psi}_0^b+\left(\beta_1^f-\frac{\beta_2^f\beta_1^b}{\beta_2^b}\right)\tilde{\psi}_1^b+\frac{\beta_2^f\beta_0^f}{\beta_2^b}\tilde{\psi}_0^f+\frac{\beta_2^f\beta_1^f}{\beta_2^b}\tilde{\psi}_1^f-i\frac{\beta_2^f}{\beta_2^b}\partial_+\tilde{\psi}_1^b.
\end{equation}

So, collecting all the coefficients of $\tilde{\psi}_1^f$, we get
\begin{equation}
\left(1-\frac{\beta_2^f\beta_1^f}{\beta_2^b}\right)\tilde{\psi}_1^f=\left(\beta_0^f-\frac{\beta_2^f\beta_0^b}{\beta_2^b}\right)\tilde{\psi}_0^b+\left(\beta_1^f-\frac{\beta_2^f\beta_1^b}{\beta_2^b}\right)\tilde{\psi}_1^b+\frac{\beta_2^f\beta_0^f}{\beta_2^b}\tilde{\psi}_0^f-i\frac{\beta_2^f}{\beta_2^b}\partial_+\tilde{\psi}_1^b.
\end{equation}
Next, we multiply this expression by  $1+\frac{\beta_2^f\beta_1^f}{\beta_2^b}$ and this gives us an explicit expression for $\tilde{\psi}_1^f$ which takes the form:
\begin{equation}
\tilde{\psi}_1^f=\left(\beta_0^f-\frac{\beta_2^f\beta_0^b}{\beta_2^b}+\frac{\beta_2^f\beta_1^f\beta_0^f}{\beta_2^b}\right)\tilde{\psi}_0^b+\left(\beta_1^f-\frac{\beta_2^f\beta_1^b}{\beta_2^b}\right)\tilde{\psi}_1^b+\frac{\beta_2^f\beta_0^f}{\beta_2^b}\tilde{\psi}_0^f-i\frac{\beta_2^f}{\beta_2^b}\partial_+\tilde{\psi}_1^b.
\end{equation}

To go further we insert this expression into the equation (\ref{psi1b}) for $\tilde{\psi}_1^b$. This leads to a complicated expression which we rewrite by collecting all the terms involving $\tilde{\psi}_1^b$.  We get
\begin{eqnarray}
\left\{1-\frac{\alpha_1^f}{\alpha_1^b}\left(\beta_1^f-\frac{\beta_2^f\beta_1^b}{\beta_2^b}\right)\right\}\tilde{\psi}_1^b+i\frac{\alpha_1^f\beta_2^f}{\alpha_1^b\beta_2^b}\partial_+\tilde{\psi}_1^b=\frac{\alpha_1^f\beta_2^f\beta_0^f}{\alpha_1^b\beta_2^b}\tilde{\psi}_0^f-\frac{i}{\alpha_1^b}\left(1+\frac{\alpha_0^f\alpha_1^f}{\alpha_1^b}\right)\partial_+\tilde{\psi}_0^b\nonumber\\
+\tilde{\psi}_0^b\left\{-\frac{\alpha_0^b}{\alpha_1^b}\left(1+\frac{\alpha_0^f\alpha_1^f}{\alpha_1^b}\right)+\frac{\alpha_1^f}{\alpha_1^b}\left(\beta_0^f-\frac{\beta_2^f\beta_0^b}{\beta_2^b}+\frac{\beta_2^f\beta_1^f\beta_0^f}{\beta_2^b}\right)\right\}.
\end{eqnarray}

To solve for $\tilde{\psi}_1^b$, we proceed in two steps. The first one involves  multiplying the equation above by the conjugate of the coefficient of $\tilde{\psi}_1^b$:
\begin{equation}
1+\frac{\alpha_1^f}{\alpha_1^b}\left(\beta_1^f-\frac{\beta_2^f\beta_1^b}{\beta_2^b}\right).
\end{equation}
This gives us:
\begin{eqnarray}
\left(1+i\frac{\alpha_1^f\beta_2^f}{\alpha_1^b\beta_2^b}\partial_+\right)\tilde{\psi}_1^b=\frac{\alpha_1^f\beta_2^f\beta_0^f}{\alpha_1^b\beta_2^b}\tilde{\psi}_0^f-\frac{i}{\alpha_1^b}\left(1+\frac{\alpha_0^f\alpha_1^f}{\alpha_1^b}+\frac{\alpha_1^f}{\alpha_1^b}\left(\beta_1^f-\frac{\beta_2^f\beta_1^b}{\beta_2^b}\right)\right)\partial_+\tilde{\psi}_0^b\nonumber\\
+\left\{-\frac{\alpha_0^b}{\alpha_1^b}\left(1+\frac{\alpha_0^f\alpha_1^f}{\alpha_1^b}\right)+\frac{\alpha_1^f}{\alpha_1^b}\left(\beta_0^f-\frac{\beta_2^f\beta_0^b}{\beta_2^b}+\frac{\beta_2^f\beta_1^f\beta_0^f}{\beta_2^b}\right)-\frac{\alpha_0^b\alpha_1^f}{(\alpha_1^b)^2}\left(\beta_1^f-\frac{\beta_2^f\beta_1^b}{\beta_2^b}\right)\right\}\tilde{\psi}_0^b.
\end{eqnarray}
Next we use the explicit expression for $\tilde{\psi}_0^f$ given above to get
\begin{equation}
\left(1+i\frac{\alpha_1^f\beta_2^f}{\alpha_1^b\beta_2^b}\partial_+\right)\tilde{\psi}_1^b=A_0\tilde{\psi}_0^b+A_1\partial_+\tilde{\psi}_0^b,
\end{equation}
where the bosonic functions $A_0$ and $A_1$ are given by expressions (\ref{A0}) and (\ref{A1}). The second and final step involves the application of the inverse of the operator acting on $\tilde{\psi}_1^b$ which is  given by
\begin{equation}
\left(1+i\frac{\alpha_1^f\beta_2^f}{\alpha_1^b\beta_2^b}\partial_+\right)^{-1}=1-i\frac{\alpha_1^f\beta_2^f}{\alpha_1^b\beta_2^b}\partial_+.
\end{equation}
This result follows directly from the fact that $(\alpha_1^f)^2=(\beta_2^f)^2=0$. This leads to the expression for $\tilde{\psi}_1^b$ given in (\ref{psi1bnew}).

\end{document}